\documentclass[useAMS,usenatbib]{mn2e}

\newcommand{\slope}{\alpha}
\newcommand{\slpar}{\lambda}
\newcommand{\br}{{\bf r}}
\newcommand{\Di}{D^{i}}
\newcommand{\Wi}{W^{i}}
\newcommand{\gi}{g^{i}}
\newcommand{\qi}{q^{i}}

\usepackage{graphicx}

\title[Cosmic magnification: nulling the intrinsic clustering signal]{Cosmic magnification: nulling intrinsic clustering}
\author[Alan F. Heavens and Benjamin Joachimi]{Alan F. Heavens$^{1}$\thanks{E-mail:
afh@roe.ac.uk; bj@roe.ac.uk} and Benjamin Joachimi$^{1}$\\
$^{1}$ SUPA, Institute for Astronomy, University of Edinburgh, Blackford Hill, Edinburgh EH9 3HJ, U.K. }
\begin{document}

\date{}

\pagerange{\pageref{firstpage}--\pageref{lastpage}} \pubyear{2010}

\maketitle

\label{firstpage}

\begin{abstract}
We investigate the extent to which the pure magnification effect of gravitational lensing can be extracted from galaxy clustering statistics, by a nulling method which aims to eliminate terms arising from the intrinsic clustering of galaxies.  The aim is to leave statistics which are free from the uncertainties of galaxy bias.    We find that nulling can be done effectively, leaving data which are relatively insensitive to uncertainties in galaxy bias and its evolution, leading to cosmological parameter estimation which is effectively unbiased.   This advantage comes at the expense of increased statistical errors, which are in some cases large, but it offers a robust alternative analysis method to cosmic shear for cosmological imaging surveys designed for weak lensing studies, or to full modelling of the clustering signal including magnification effects.
\end{abstract}


\section{Introduction}

Cosmic shear has long been recognised as a potentially very powerful tool for determining the properties of the Universe.  Its sensitivity to cosmological parameters and its 
clean dependence via gravity on the mass distribution make it an attractive option for determining, amongst other things, the properties of Dark Energy, and it also opens up the prospect of testing beyond-Einstein gravity models.  For recent reviews and summaries, see \citet{MVWH,HoekstraJain08,CosmicVision,Albrecht06}.  With the advent of very large surveys directed towards large-scale weak lensing effects, such as Pan-STARRS1, the Dark Energy Survey, and in the future, LSST, Euclid and WFIRST, focus has shifted towards the systematic errors rather than the raw statistical power of weak lensing.   For cosmic shear, there are challenges related to the precise measurement of galaxy shapes (e.g. \citealt{KTH08,AR08,VB10,Bernstein10,GREAT10}), and physical effects such as intrinsic alignments (e.g. \citealt{HRH,croft00,crittenden01,catelan01,jing02,mackey02,HS04,bridle07a,schneiderm09}) which can mimic to some degree the effects of weak lensing.  These effects will increase the errors on recovered parameters, so it is timely to investigate magnification (or amplification) - another feature of weak gravitational lensing, which, although not as statistically powerful as cosmic shear, provides independent cosmological information.  Given that the effects of changing Dark Energy properties and changing the gravity law are rather subtle, it is highly desirable to have independent techniques which are not subject to the same systematic errors.  

Weak lensing induces small changes to object sizes, which, through the conservation of surface brightness, leads to flux changes which alter the number density of galaxies when a flux limited survey is considered.  As a result of this amplification, in concert with the changes in positions, the clustering pattern of galaxies is altered.   In regions where there is magnification (as opposed to de-magnification), the number density is increased or decreased depending on the slope of the number counts, as it depends on how many sources are promoted into the survey by the lensing, and \citet{VW10} has shown how magnification and shear can be used together to help cosmological parameter estimation.  Recent studies \citep{Scranton05,Hil09a,Menard} have demonstrated the magnification effect with cosmological surveys, and \cite{ZP05,ZP06} have shown how flux information may be used to isolate lensing magnification.

The main complication which has to be dealt with is that there is a strong intrinsic clustering of galaxies, which contaminates the clustering signal due to weak lensing.  Indeed, to call it a contaminant is a misnomer, as it is much stronger than the lensing signal.  The intrinsic clustering of galaxies is not as easy to model theoretically as the lensing signal, as the former depends on galaxy formation efficiency (normally distilled into {\em galaxy bias}), whereas the lensing signal depends only on the matter distribution.  One approach is to model the entire clustering system, including lensing and bias, and this has been the approach of \citet{Bernstein, Yoo, JB10}.  However, since the main advantage of lensing is its independence (at least in its dominant effect) on complex astrophysics which may be difficult to determine with precision,  it is appealing to consider an alternate strategy, where one removes the intrinsic clustering part of the signal, leaving, ideally, the pure lensing signal.  If this can be achieved, then confrontation with theory becomes much more straightforward and robust.  The purpose of this paper is to investigate to what extent this can be achieved, using a `nulling' technique similar to that which has been proposed to deal with intrinsic alignments in cosmic shear.   Nulling can be achieved because we know (for a given cosmology) how the source of the magnification signal depends on redshift, and we can exploit this to find combinations of correlations which do not include intrinsic clustering terms.

The problems of dealing with intrinsic alignments in cosmic shear and intrinsic clustering in cosmic magnification are in some respects similar, but in others different.  In both cases the observable quantity (the ellipticity in the case of shear, the number density in the case of magnification) depends to first order on the addition of two components, so any quadratic quantities such as the correlation function or power spectrum of the observables have four terms.  The main differences are that in the cosmic shear case, the contaminants are relatively small, but poorly-known theoretically, whereas in the magnification case, the contaminating terms are large, but in principle measurable.

The structure of the paper is as follows.  In section \ref{theory} we review the theory of cosmic magnification.  In section 3 we develop the nulling theory, and in section 4 we present results, with conclusions given in section 5.

\section{Cosmic Magnification and Amplification}
\label{theory}

The distortion of light bundles leads to a mapping from the source plane to the image plane which is described by a symmetric amplification matrix which is decomposed into a convergence $\kappa$, and two shear terms $\gamma_1$ and $\gamma_2$, which describe distortions along the coordinate axes and at $45^\circ$ respectively:
\[
A = \left(
\begin{array}{cc}
1-\kappa + \gamma_1 & -\gamma_2 \\ -\gamma_2 & 1-\kappa-\gamma_1\\
\end{array}
\right).
\]
Size changes (and hence flux changes, by Liouville's theorem)  are related to the inverse of the Jacobian of the transformation, which to linear order is 
\[
\mu \simeq 1 + 2\kappa + \mathcal{O}(\kappa^2).
\]
This is the magnification or amplification.  Let the unmagnified average number density at position $\br$ above a flux limit $f$ be $\bar n(>f,\br)$, and we assume that near the flux limit of a survey it has a slope $\slope$, i.e. 
\[
\bar n(>f,\br) \propto f^{-\slope}
\]
then the expected observed galaxy counts, neglecting intrinsic clustering, for given magnification $\mu(\br)$  will be
\[
n(>f,\br) = \bar n(>f,r)\mu(\br)^{\slope -1}.
\]
 The solid angle elements are also magnified, thus lowering the number density and resulting in the $\mu^{-1}$ factor \citep{Narayan89}.   Note that the slope may be dependent on $\br$ (only via $r\equiv |\br|$ if the survey selection is isotropic within the observed area), and that the power-law need not extend over a wide range of $f$; it is the local slope at the flux limit which is required.    If we include intrinsic clustering of the sources, characterised by a fractional overdensity $\delta_g$, then the observed number density of galaxies at position $\br$ is given by (dropping the $>f$)
\[
n({\bf r}) = \bar n(r)\mu(\br)^{\slope-1}\left[1+\delta_g(\br)\right]
\]
where $\delta_g$ is the intrinsic fractional overdensity of galaxies.  If we define $b$ as the bias parameter, here in real space, so $\delta_g \equiv b \delta$, where $\delta$ is the fractional mass overdensity, then linearising gives
\[
n(\br) = \bar n(r)\left[1+ \slpar \kappa + b \delta(\br)\right],
\]
where $\slpar\equiv 2(\slope-1)$ may be a function of $r$, as may $b$, through time-evolution.   $\slpar$ should be relatively easy to measure from sufficiently deep photometric catalogues, so we initially regard it as fixed and known.  The convergence is 
\[
\kappa(\br) = A\int_0^r dr' \,F_K(r,r')\,\delta(\br')
\]
where $F_K(r,r') = r'(r-r')/[r a(r')]$ for a flat Universe (for $r>r'$, 0 otherwise), and $A=3H_0^2\Omega_m/(2c^2)$.  We will assume flatness, but this can be easily generalised. Thus since $\kappa$ is linear in $\delta$, we have to linear order that the fractional number overdensity $\Delta \equiv n/\bar n-1$ is:
\[
\Delta(\br) = \slpar(r)\kappa(\br)+b(r)\delta(\br).
\]
Let us consider the two-point statistics of $\Delta$, for a tomographic survey divided into $n_{bins}$ shells in estimated redshift (normally coming from photometric redshift estimates).  We will frame this discussion in real space, but will switch to Fourier space in section \ref{nulling}.   The two-point cross-correlation function of the number overdensity in two shells $i$ and $j$ is (with obvious notation)
\begin{eqnarray}
\langle \Delta_i \Delta_j\rangle &=& \slpar_i\slpar_j\langle \kappa_i \kappa_j \rangle + b_i b_j\langle \delta_i \delta_j\rangle\nonumber\\
&+& \slpar_i b_j \langle\kappa_i  \delta_j \rangle+  b_i\slpar_j \langle\delta_i \kappa_j\rangle.\qquad i,j = 1,\ldots n_{bins}\nonumber
\end{eqnarray}
If we wish to have a clean cosmological test, then it is the first term which is of most interest, as it does not contain galaxy bias, which is not precisely known.  If we are to use it, then we need to remove or model the other terms.  In this paper we choose the former strategy, which can be achieved by nulling out the remaining terms.   This is very similar to what has been proposed by \citet{JS08} and \citet{JS09} for removal of intrinsic alignment (IA) terms in cosmic shear, where the second term is analogous to the so-called II term, and the cross-terms are GI terms in cosmic shear (and only one survives in that case).   We turn to how to do this in the next section.

There are two main differences between magnification nulling and IA.  One is the magnitude of the effect.  GI is generally subdominant to the desired GG shear signal, whereas here the $\kappa\delta$ terms far exceed the bias-independent $\kappa\kappa$ term.  See the top panels of Fig. 1.  The other difference is that we have rather poor theoretical understanding of the GI term, whereas we can in principle make some measurements of bias of galaxies.

\section{Nulling}
\label{nulling}

As with the shear nulling method of Joachimi \& Schneider (2008), we cross-correlate the overdensity of galaxies in a tomographic bin labelled by $i$ with a weighted average of the overdensities in all other tomographic bins.  Writing this average as $D^{i}(\btheta)$  in the continuum limit, where $\btheta$ is an angular position on the sky,
\[
\Di(\btheta) = \int_0^\infty dr \, \Wi(r) \Delta({\br}).
\]
If we choose $\Wi(r)$ to be zero for $r$ close to $r_i$ (the centre of the tomographic bin $i$), then the intrinsic clustering term (a product of $b\delta$ terms) will be small. This is analogous to the downweighting \citep{HH03} or removal of \citep{KS02,KS03} close pairs to remove the II term in cosmic shear.   

We also want to zero out the cross terms ($\kappa \delta$ terms).  One cross term is
\[
A b(r_i)\delta(\br_i)  \int_0^\infty dr \, \Wi(r) \slpar(r) \int_0^r dr' F_K(r,r') \delta({\br'}),
\label{weightconstraint}
\]
and we choose $\Wi$ such that this is zero.  Reversing the integration order (and dividing by $A$),
\begin{eqnarray}
0 &=& b(r_i)\delta(\br_i)  \int_0^\infty dr' \int_{r'}^\infty dr \, \Wi(r) \slpar(r)\frac{(r-r')r'}{r a(r')}\delta(\br')\nonumber\\
& = & b(r_i)\delta(\br_i)\int_0^\infty dr \frac{r}{a(r)}\gi(r)\delta(\br)
\label{cross}
\end{eqnarray}
where
\begin{equation}
\gi(r) \equiv \int_r^\infty dr' \Wi(r') \slpar(r')\frac{(r'-r)}{r'}. \label{constraint}
\end{equation}
If we choose $\gi(r)=0$ on scales where the density correlations are significant (i.e. at $r \simeq r_i$), then the expectation value of equation $\ref{cross}$ will be nearly zero.  With $r_i$ at the median redshift of the bin, we choose
\[
\gi(r_i) = 0.
\]
Since $\kappa(r_i)$ is influenced only by $r<r_i$, we null the other cross term by simply requiring
\[
\Wi(r)=0; \qquad r<r_i.
\]
There are many weight vectors which will satisfy this equation, so for maximum statistical power, we follow \citet{JS09} and include all orthogonal weight vectors which satisfy the constraint.  Indeed the weights are identical to the intrinsic alignment case if the slope of the number counts does not vary with redshift,  but is modified if it does.    For the details of how this is done, the reader is directed to \citet{JS09}.  Note that the weights require the choice of a fiducial cosmology, and the weights must not be changed as a search of parameter space is undertaken.  Clearly, if the choice is very poor, the nulling will be suboptimal, but there would be no difficulty in iterating the process, where the best-fit cosmology is used to define optimal nulling weights, and the process repeated.  In this paper, we assume the correct cosmology for the weights, for simplicity.

To compute the weighted cross-power spectrum, we note that we are cross-correlating two fields 
\begin{eqnarray}
u_0^{i}(\btheta) &\equiv& \slpar(r_i)\kappa(r_i)+b(r_i)\bar\delta(r_i)\\\nonumber
u_s^{i}(\btheta) & \equiv & \int_{r_i}^\infty dr W_s^{i}(r)\left[\slpar(r)\kappa(\br)+b(r)\delta(\br)\right]\qquad s>0,
\end{eqnarray}
where the first term averages the overdensity over the shell.  In the second term, we allow for a variety of weighting functions $s=1,\ldots n_{bins}-1$, as there are $n_{bins}-1$ orthogonal weighting functions which satisfy the constraint equation  \ref{constraint}.

We can write these in the form
\[
u_s^{i} = \int_0^\infty dr \, \qi_{s}(r) \delta(\br)
\]
where
\begin{eqnarray}
\qi_0(r) &=& A\slpar(r)\frac{(r_i-r)r}{r_i a(r)} + b(r)T(r-r_i)\\\nonumber
\qi_s(r) &=& A\gi(r)\frac{r}{a(r)} + b(r) \Wi(r)\qquad s>0
\end{eqnarray}
where $T$ is roughly a top-hat of unit area centred on the bin centre $r_i$.  This can be refined to reflect the averaging over the bin width with a varying number density.

For such fields, the Limber approximation (e.g. \citealt{BS01}, section 2.84) gives the cross-power spectrum as
\[
P^{ii}_{0s,\ell} = \int_0^\infty dr \frac{q^{i}_0(r)q^{i}_s(r)}{r^2} P(\ell/r;r)
\]
where the last argument of the matter power spectrum $P$ indicates its time-dependence.

\subsection{Fisher analysis}

The sensitivity to cosmological parameters comes from the matter power spectrum and its evolution with time, and the distance-redshift relation
\[
r(z) = c\int_0^z\frac{dz'}{H(z')}
\]
where $H(z) = H_0\left[(1-\Omega)(1+z)^2   +\Omega_{DE}   + \Omega_m(1+z)^3\right]^{1/2}$ if the Dark Energy has present density parameter $\Omega_{DE}$. $\Omega_m$ and $\Omega$ are the present matter and total density contributions, and we ignore radiation.  Note that later we allow the Dark Energy density to evolve, so this equation is modified in the standard way.

For this initial study, we compute the covariance of power spectrum estimates  assuming gaussian statistics, but note that this would need to be estimated from numerical simulations (e.g. \citealt{Kiessling10}) for real applications, and would include non-zero cross terms.  The covariance of powers averaged in bins of width $\Delta \ell$ is 
\[
C_{ij}(\ell) \equiv Cov(P_{0s,\ell}^{ii},P_{0s',\ell}^{jj}) = \frac{2\pi}{\Delta\Omega\, \ell \Delta\ell} (P^{ij}_{00,\ell}P^{ij}_{ss',\ell}+P^{ij}_{0s',\ell}P^{ij}_{s0,\ell})
\]
where $\Delta\Omega$ is the solid angle of the survey, and 
\[
P^{ij}_{ss',\ell} \equiv \int_0^\infty  \frac{dr}{r^2} q^{i}_s(r) q^{j}_{s'}(r) P(\ell/r;r) + {\rm shot\ noise}
\]
is a function of the bins $i$ and $j$.   For $\ell\ne\ell'$ the covariance is zero if we assume all-sky coverage.  For practical cases we use the usual scaling with the sky fraction $f_{\rm sky}$ if, as is the case here, most of the signal comes from relatively small scales.

Finally, we compute the Fisher matrix \citep{TTH}  for the cosmological parameters $\theta_\alpha$: 
\begin{eqnarray}
F_{\alpha\beta} &=&   \sum_{\ell\ bins} F_{\alpha\beta}(\ell),\nonumber\\
F_{\alpha\beta}(\ell) &=& \frac{\partial P_a}{\partial\theta_\alpha} C_{ab}^{-1}(\ell) \frac{\partial P_b}{\partial\theta_\beta},
\label{Fisher}
\end{eqnarray}
and the summation convention is assumed for $a,b$, which are indices which run over all the $P^{ii}_{0s}$ modes.

The Fisher matrix also allows us to compute analytically the parameter biases arising from incorrect parametrisation of the data, assuming a gaussian likelihood surface  (see e.g. \citealt{Knox98,Taylor07}):
\[
\delta\theta_\alpha = -[F_{\theta\theta}]_{\alpha\gamma}^{-1}(F_{\theta\psi})_{\gamma\beta}\delta\psi_\beta
\]
where $\delta\psi_\beta$ is the error made in the fixed model parameter $\theta_\beta$, and $F_{\theta\psi}$ is a pseudo-Fisher matrix of the same form as equation \ref{Fisher}, but with one $\theta$ replaced by $\psi$.

\section{Results}

\begin{figure*}
\centering
\includegraphics[scale=.6,angle=0]{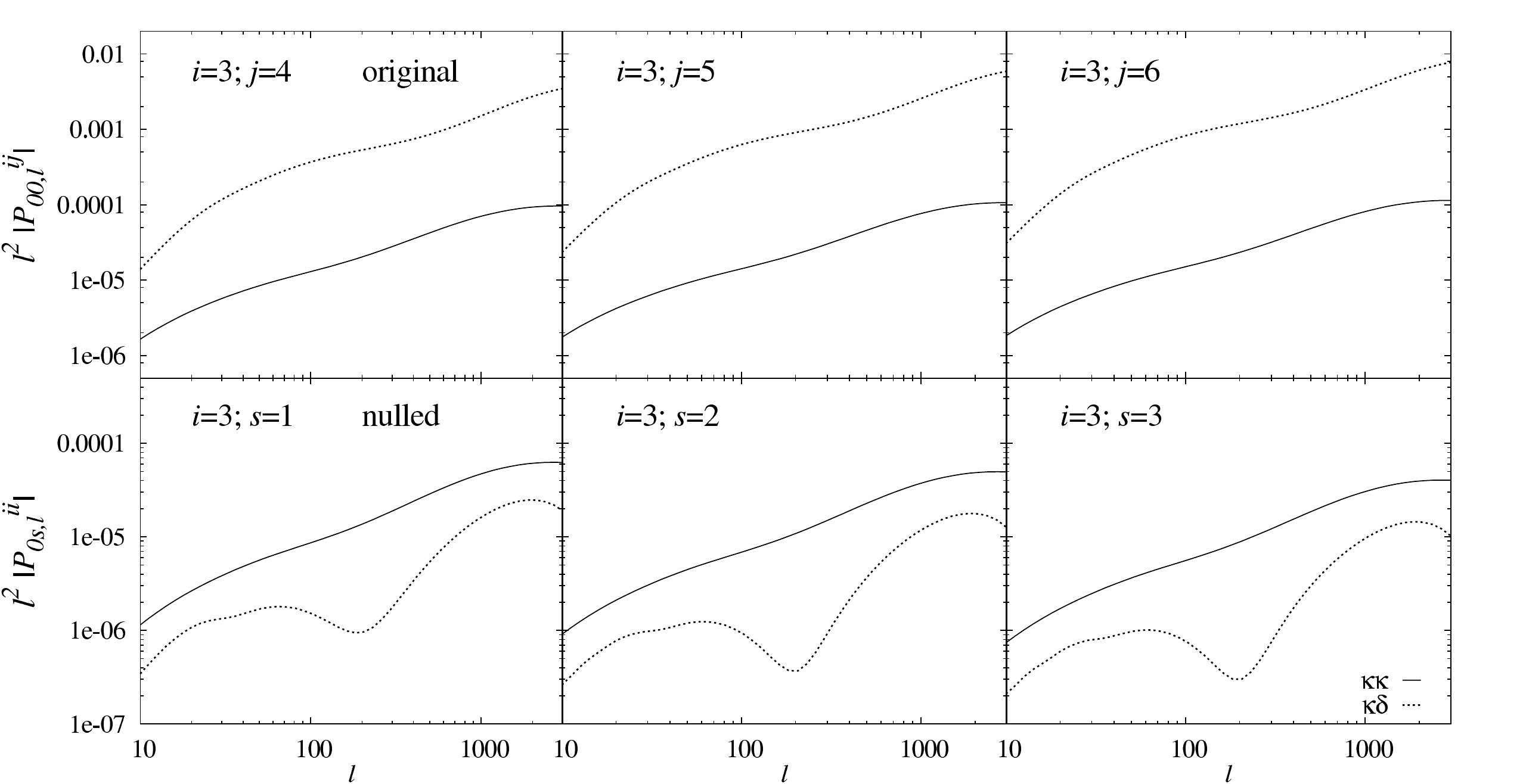}
\caption{Nulling performance for an example set of power spectra for foreground bin $i=3$. Magnification correlations ($\kappa\kappa$ terms) are given as solid curves, magnification-clustering correlations ($\kappa\delta$) as dotted lines. The top panels show from left to right the cross  power spectra of bin 3 ($z\simeq 0.46$) tomographic bins $j=4,5,6$ (with central redshifts $z=0.53, \ 0.59$ and $0.65$. 
The bottom panels show nulled power spectra for the first three orthogonal weight vectors  $s=1,2,3$ (note that in the variant of nulling employed in this work no particular ordering is associated with $s$,  which each represent a linear combinations of signals from different tomographic bins.  See text for more details).}
\label{fig:nullexample}
\end{figure*}

The situation which we investigate is as follows.  The galaxy bias is scale-independent, but redshift-dependent:
\[
b(z) = 1 + mz + \Delta b
\]
with $m=0.1$ and $\Delta b=0.2$.  For the analysis, we incorrectly model the bias as a constant, independent of $z$, and treat it as a parameter in the modelling, with a gaussian prior of width 0.1, centred on unity.   We also have 6 cosmological parameters (scalar spectral index, density parameter in matter and baryons, Dark Energy equation of state parameter $w_0=p/(\rho c^2)$, fluctuation amplitude and Hubble parameter, with fiducial values $n_s=1$, $\Omega_m=0.25$, $\Omega_b=0.05$, $w_0=-1$, $\sigma_8=0.8$, $h=0.7$.  We take a number count slope $\alpha=1.5$, independent of redshift.  Note that this is a simplification to illustrate the method; in practice $\alpha$ depends strongly on magnitude limits and redshift. See, for example, the discussion in Appendix C of \cite{JB10}.

For the survey, we assume a Euclid-like basic survey design of 20,000 square degrees, but initially we will ignore photometric redshift errors for illustration.   We assume 35 galaxies per square arcminute, with a redshift distribution $n(z) \propto z^2 \exp[-(z/z_0)^{1.5}]$, and a median redshift of 0.9.  The distribution is truncated at $z=3$, and 20 redshift bins are used, chosen such that the number in each bin is constant, with 50 angular power spectra computed on a logarithmic grid between $l=10$ and $3000$.   In all cases, the auto-power spectra are not considered, to suppress the $\delta\delta$ term.

In Fig.~\ref{fig:nullexample} we show in the lower panels the effect of nulling on the cross-power spectra.  We see that if we have excellent redshifts for the galaxies, then the intrinsic clustering term $\langle \kappa_i \delta_s\rangle$ can be nulled out very effectively, reducing its relative amplitude by nearly two orders of magnitude so that it is well below the $\langle \kappa_i \kappa_s\rangle$ term (here we use a compact notation, meaning the cross-power of the $\kappa$ component of the overdensity in bin $i$ with the relevant component of one of the weightings $s$ associated with that bin).  The nulling is not perfect, because of the finite width of the bins: the nulling works at the median redshift of the bin, but is not perfect away from the bin centre.    We see that we can reduce the contamination to a level generally below the  $\langle \kappa_i \kappa_s\rangle$ term, but it is still non-negligible.  This figure suggests that some modelling of the bias is still needed, however the accuracy required is reduced.

\subsection{Modelling vs nulling}

The two general approaches we can take are to model the data, including a bias model, or to attempt to null out the terms depending on bias, and to model the remaining terms, which depend only rather cleanly on cosmological parameters.  Our results of the effectiveness of nulling indicate that the latter approach is too ambitious, as the residual $\kappa\delta$ signal is not small enough to be ignored, and it must be modelled.  Having said this, its sensitivity to bias is much smaller than without nulling, so there are some advantages to this approach.   What we show in this section is that if the modelling of bias (and its evolution) is imperfect, then modelling the complete clustering signal can lead to large biases in cosmological parameter estimates (in terms of the statistical error), whereas with nulling, the parameter biases are typically much less than the statistical errors. On the other hand, the statistical errors with nulling may be considerably larger than in the non-nulled case, so nulling gives us a rather conservative, but robust analysis.

We perform various analyses.   In Version 1 (nulled), we perform nulling to remove the $\kappa\delta$ terms as far as possible.  Here we assume that the nulling is perfect, and there is no signal left except for the pure $\kappa\kappa$ term.   Because of the finite width of the redshift bins, the nulling is not perfect, and leaves a residual $\kappa\delta$ term, which leads to some bias in the parameter estimates, some of which remain unacceptably large, and very large statistical errors.   Thus we conclude that even when the photometric redshifts are assumed to be known exactly, one cannot ignore the residual $\kappa\delta$ contamination in the nulling, and it has to be modelled.

\begin{table*}
\begin{minipage}{120mm}
\centering
\caption{Resulting $1\sigma$ marginalised statistical errors ($\sigma$) and parameter biases ($b$, not to be confused with galaxy bias) from Fisher matrix predictions for different analysis strategies. Version 1 corresponds to treating $\kappa\kappa$ as the signal and ignoring the contamination by the $\kappa\delta$ term, which acts as a systematic error. Version 2 corresponds to treating the sum of $\kappa\delta$ and the imperfectly nulled $\kappa\delta$ term as the signal, marginalising over the assumed constant galaxy bias in the latter contribution.  Here a systematic error arises because the bias model for the $\kappa\delta$ term is not correct, but we see that nulling very effectively reduces this systematic to a low level.}
\begin{tabular}[t]{c|cc|cc|cc|cc}
\hline\hline
parameter &  \multicolumn{2}{c|}{Version 1; nulled} & \multicolumn{2}{c|}{Version 2; standard} & \multicolumn{2}{c}{Version 2; nulled}\\
 & $\sigma$ & $b$ & $\sigma$ & $b$ & $\sigma$ & $b$\\ 
\hline
$\Omega_{\rm m}$ &  0.463 &  0.138 & 0.012 &  0.000 & 0.078 & -0.021 \\
$\sigma_8$       &  0.516 & -0.222 & 0.009 &  0.020 & 0.120 &  0.038 \\
$h$              &19.650 &  6.238 & 0.134 &  0.382 & 0.554 & -0.059 \\
$n_{\rm s}$       &  3.888 & -1.165 & 0.051 & -0.165 & 0.211 &  0.033 \\
$\Omega_{\rm b}$  & 2.324 &  0.760 & 0.011 &  0.024 & 0.030 & -0.006 \\
$w_0$            &  4.780 & -0.239 & 0.088 &  0.202 & 0.267 &  0.008 \\
\hline
\end{tabular}
\label{tab:magnullresults}
\end{minipage}
\end{table*}

\begin{figure*}
\begin{minipage}{120mm}
\centering
\includegraphics[scale=.6,angle=270]{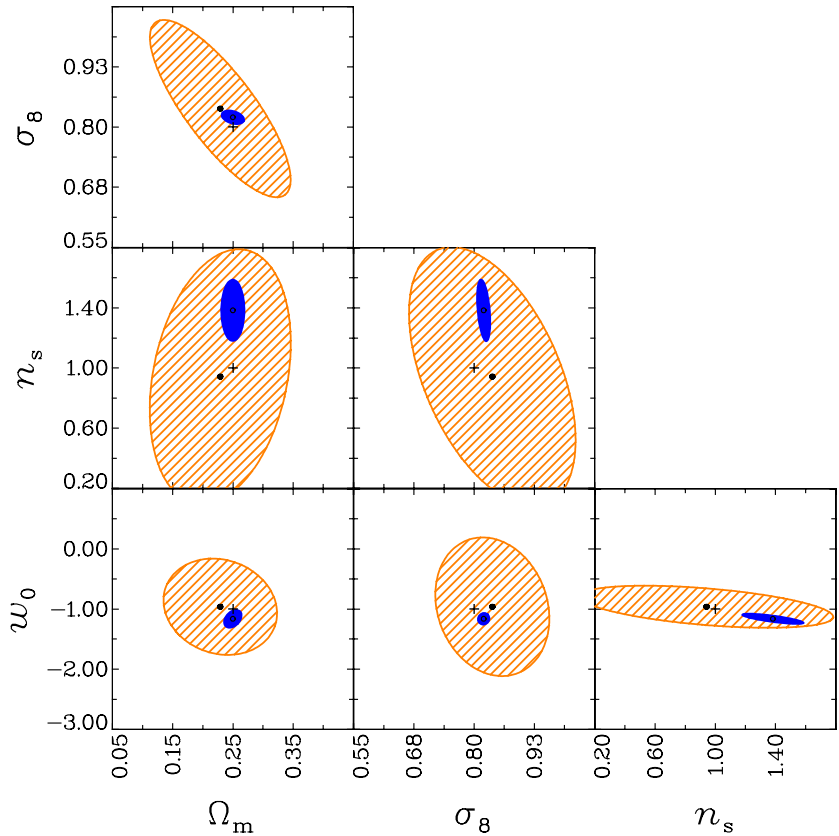}
\caption{$1\sigma$ constraints from the $\kappa\kappa$ and residual $\kappa\delta$ signals, including marginalisation over the galaxy bias. Blue filled contours correspond to marginalisation only (3rd column of Table \ref{tab:magnullresults}); orange shaded contours correspond to marginalisation + nulling (4th column of Table \ref{tab:magnullresults}). Parameters not shown are marginalised over. The cross in each panel marks the fiducial parameter values.}
\label{fig:magnullcontours}
\end{minipage}
\end{figure*}

In Version 2, we model the residual $\kappa\delta$ terms.  In the standard case, we are essentially modelling the entire signal, without attempting to reduce the $\kappa\delta$ term by nulling.   We see from Table \ref{tab:magnullresults} that for the bias model we investigate, the incorrect assumption for the bias evolution (i.e. that there is none) leads to residual biases which are typically quite large compared to the statistical errors.   The extent of the biases is dependent on how wrong our bias model assumptions are - if we had chosen the correct parametrisation for the bias evolution, we would not expect a bias in the recovered parameters.   However, it is not clear 
how well we will be able to constrain the bias.  We can use the microwave background radiation and the measured galaxy power spectrum, if we assume a gravity model, or use higher-order statistics (e.g. \cite{Verde02}), but it will probably be difficult to constrain the bias very well.
In the nulled case, we remove most of the $\kappa\delta$ signal, but we model the residual contamination.  Since the contamination (which depends on the galaxy bias) is relatively small, the error in the assumed bias evolution has a much smaller effect, and the biases in the parameter estimation are negligible in comparison with the statistical errors.   The main reason for this is that the statistical errors have grown substantially (by factors of $3-12$), but we see from the table that for most parameters, the biases are in absolute terms smaller than in the no-nulling case.  These results are illustrated in Fig. 
\ref{fig:magnullcontours} for a few of the cosmological parameters.  The nulled statistical errors (orange, shaded regions) are much larger than modelling the full signal (blue, solid regions), but the bias is typically smaller.  Note, however, that the performance of nulling compared with full modelling depends on how wrong the assumed bias model is.  If our model is poor, then nulling can fare better in absolute terms. We turn to this and related issues next.

\subsection{Sensitivity to assumptions}

\begin{figure*}
\begin{minipage}[c]{0.5\textwidth}
\centering
\includegraphics[scale=.35,angle=0]{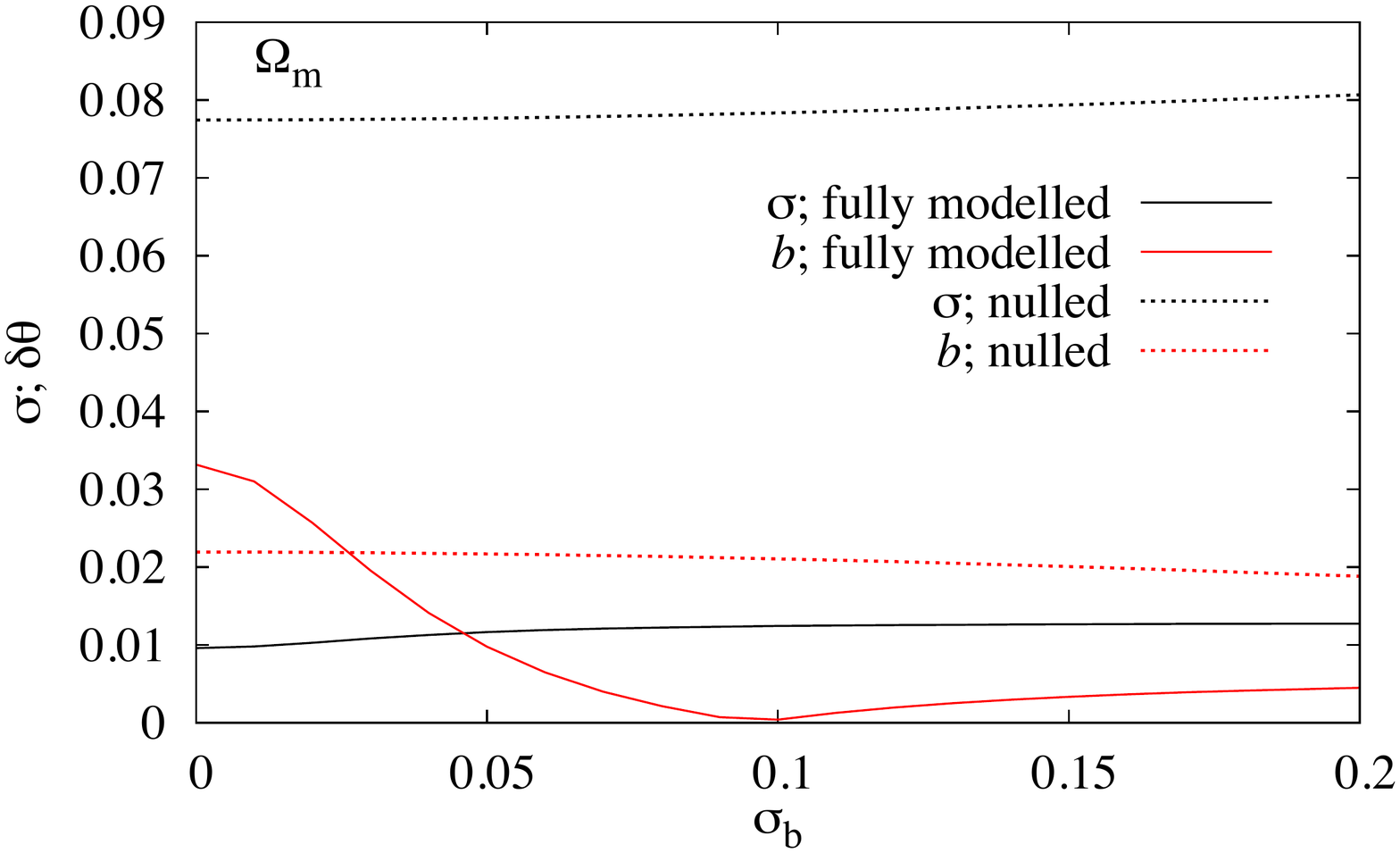}
\includegraphics[scale=.35,angle=0]{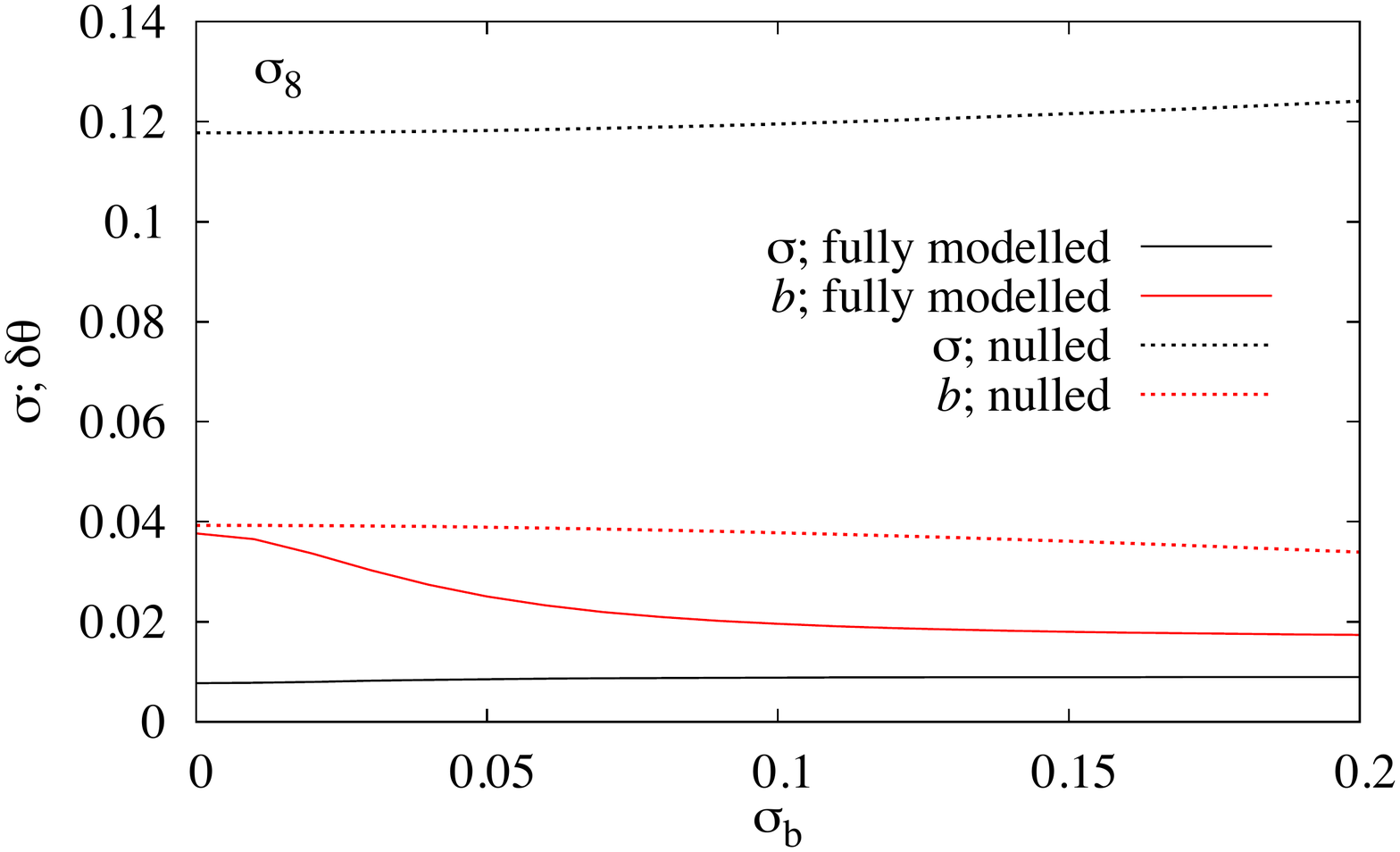}
\end{minipage}%
\begin{minipage}[c]{0.5\textwidth}
\centering
\includegraphics[scale=.35,angle=0]{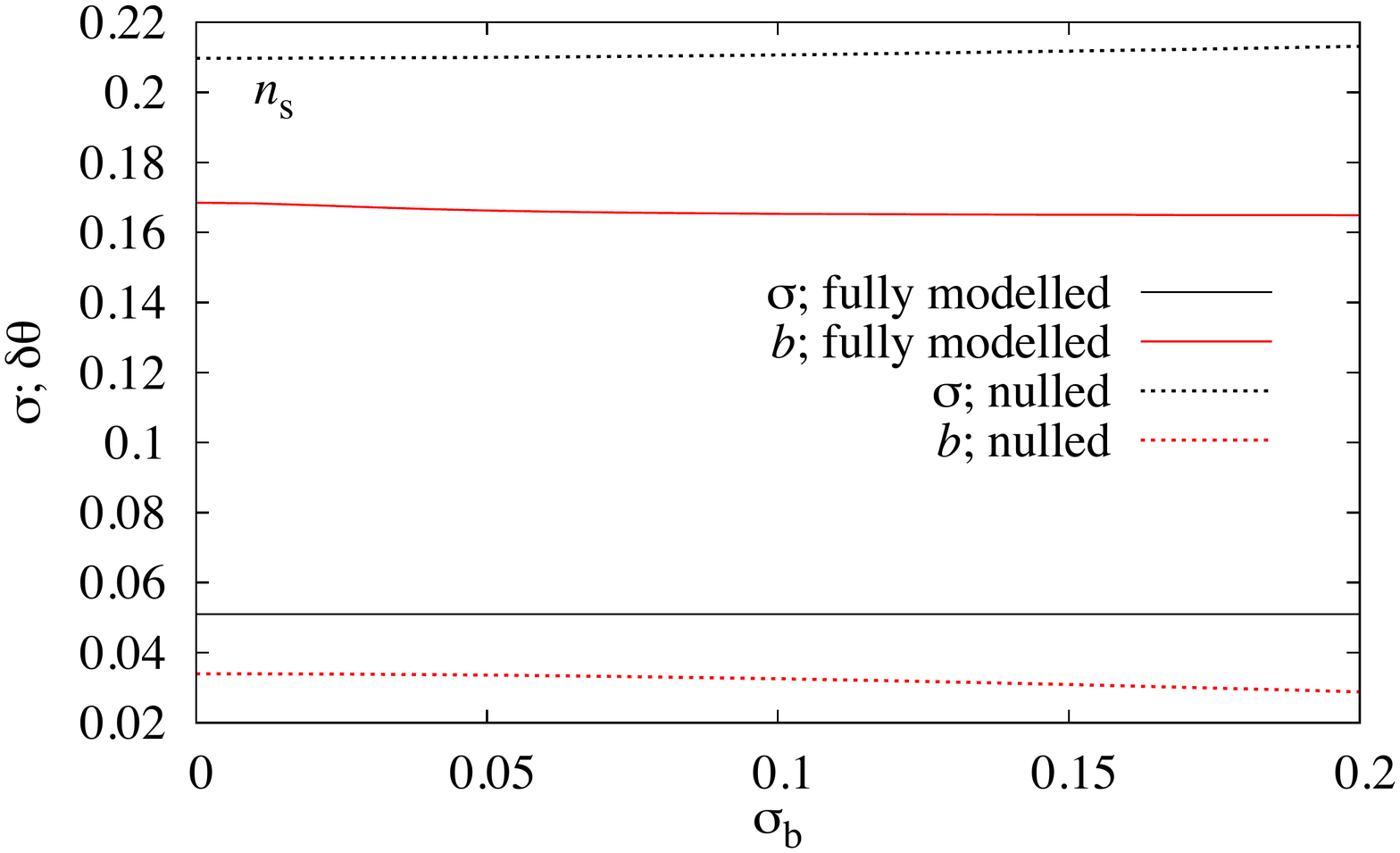}
\includegraphics[scale=.35,angle=0]{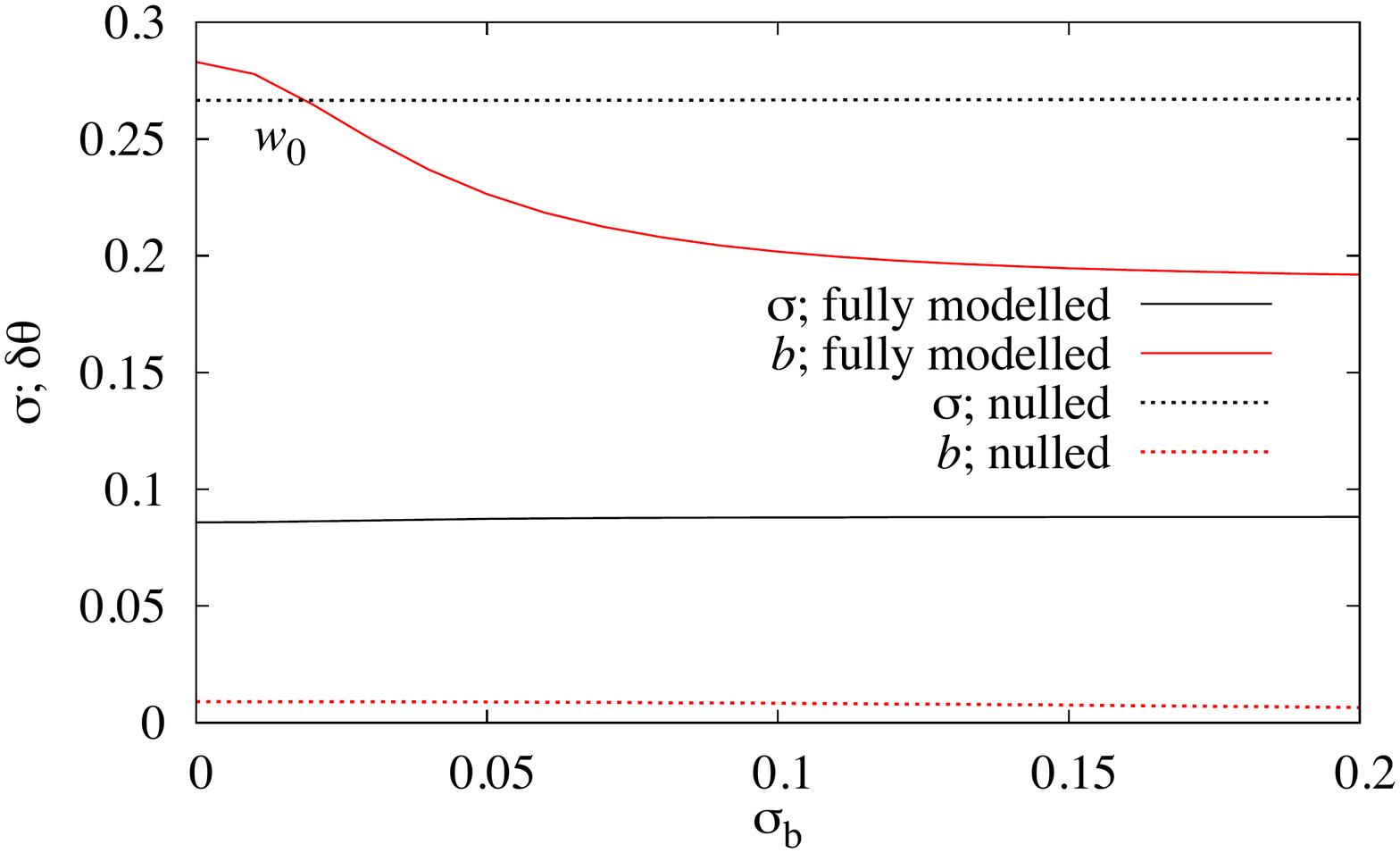}
\end{minipage}
\caption{Statistical $1\sigma$ errors (black lines) and biases (red/grey lines) for different cosmological parameters as a function of the prior on the galaxy bias parameter, keeping all other  bias model parameters at their fiducial values. Solid curves correspond to the results for the fully-modelled power spectrum analysis including marginalisation over the galaxy bias, dotted curves to the combined nulling and galaxy bias marginalisation approach.}
\label{fig:stats_prior}
\end{figure*}

In Fig. \ref{fig:stats_prior} we illustrate how sensitive the errors are to an assumed prior on the galaxy bias. The prior is assumed to be gaussian, centred on unity, with r.m.s. $\sigma_b$, labelling the $x$ axis.    Each panel illustrates the errors on a cosmological parameter, marginalised over the others.  General features are that the nulled statistical errors are always larger than the fully-modelled case.  In most cases, the nulled bias is smaller than the fully-modelled case (for $\Omega_m$ the fully-modelled bias is relatively small, but this seems to be anomalous).   The nulled bias is always negligible in comparison with the nulled statistical errors, but the opposite is the normal situation for the fully-modelled case.   This is accentuated when the prior on the galaxy bias is very small - we are essentially enforcing an incorrect model for the bias.   When we increase the width of the galaxy bias prior, it is still unable to model the bias evolution (as it assumes there is none), but at least it allows the assumed constant bias to increase to closer to the average bias of the galaxies (at redshift 1, the galaxy bias is 1.3).

\begin{figure*}
\begin{minipage}[c]{0.5\textwidth}
\centering
\includegraphics[scale=.35,angle=0]{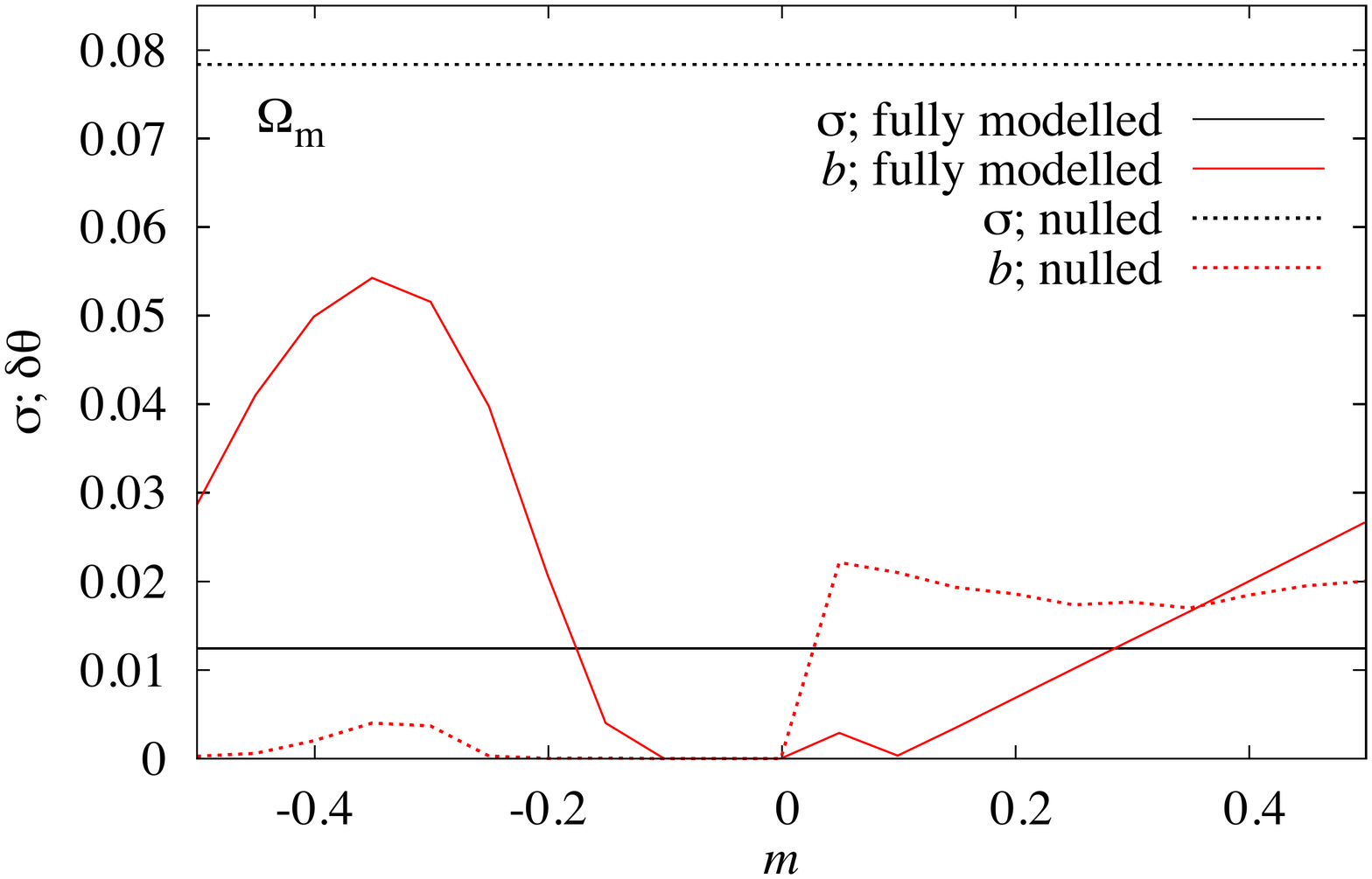}
\includegraphics[scale=.35,angle=0]{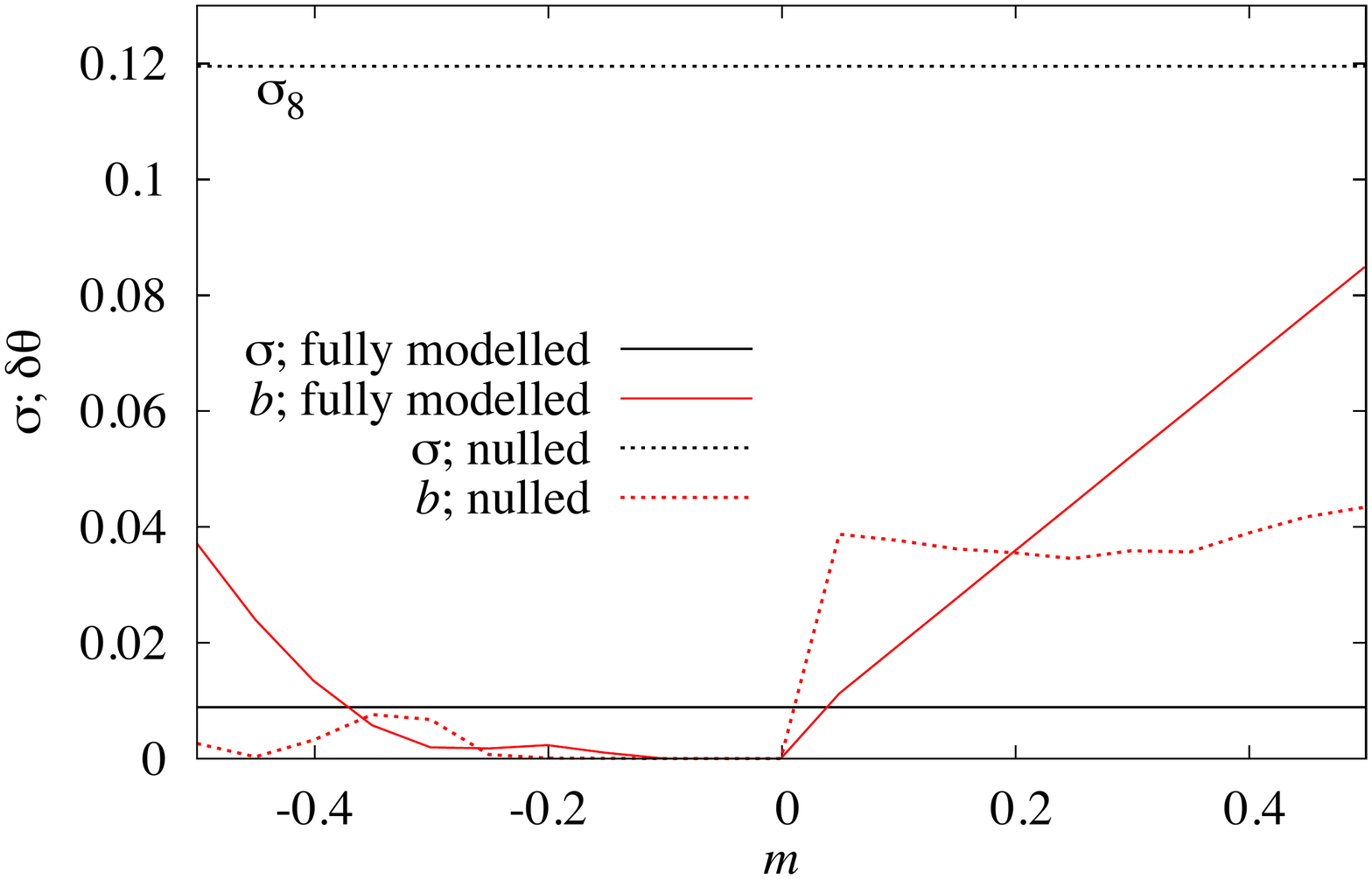}
\end{minipage}%
\begin{minipage}[c]{0.5\textwidth}
\centering
\includegraphics[scale=.35,angle=0]{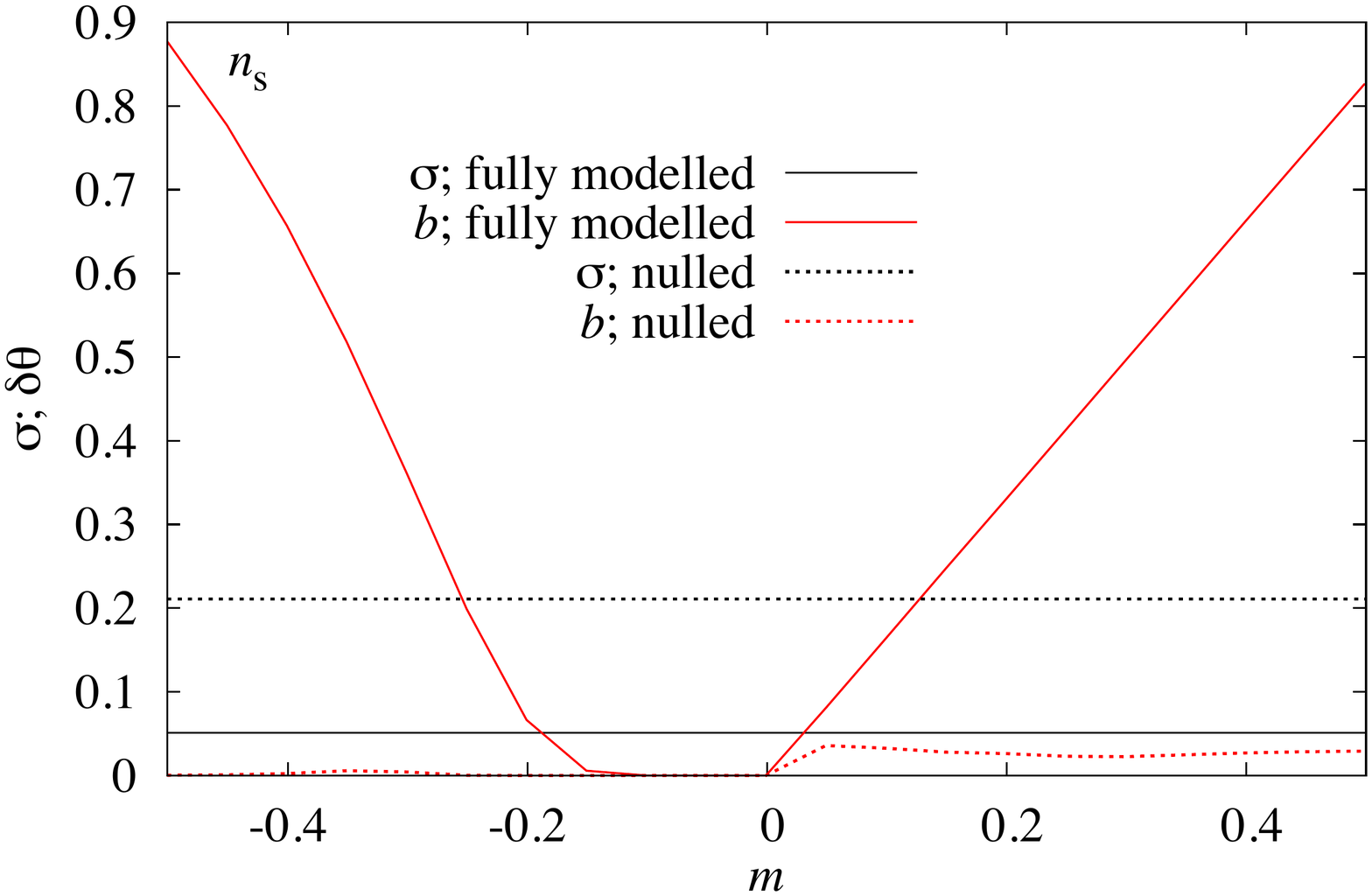}
\includegraphics[scale=.35,angle=0]{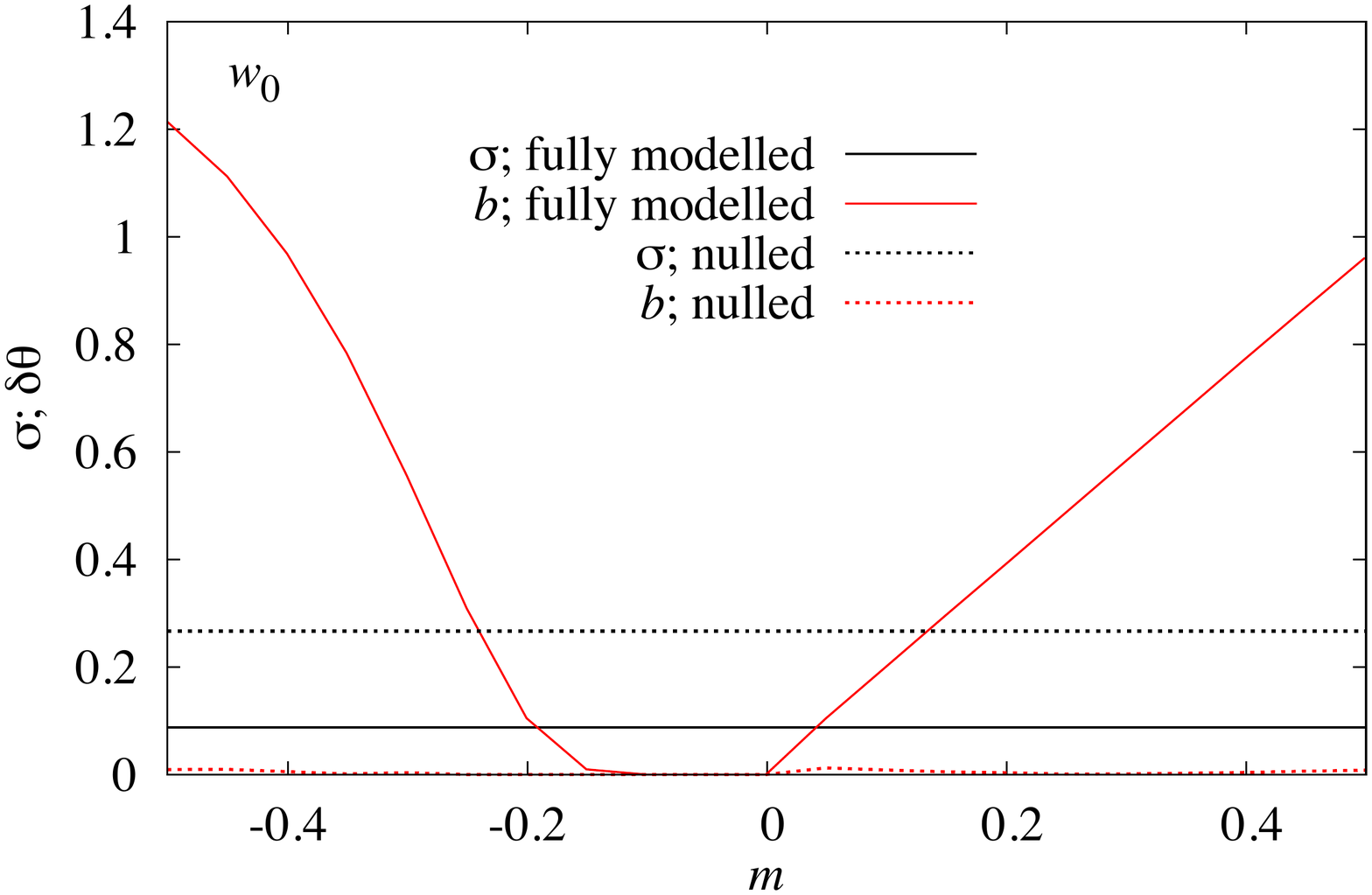}
\end{minipage}
\caption{Statistical $1\sigma$ errors (black lines) and biases (grey/red lines) for different cosmological parameters as a function of the slope $m$ of the galaxy bias model, keeping all other parameters at their fiducial values. Solid curves correspond to the results for standard analysis including marginalisation over the galaxy bias, dotted curves to the combined nulling and galaxy bias marginalisation approach.}
\label{fig:stats_slope}
\end{figure*}

In Fig. \ref{fig:stats_slope} we see the effect of varying the slope $m$ of the galaxy bias evolution model.  This illustrates a main conclusion of this study, which is that nulling is particularly effective if one makes incorrect assumptions about the model parametrisation.  Here $m$ is fixed, and not marginalised over, and all the panels in the figure show that the worse the parametrisation, the greater the biases in the parameter estimates.  The solid grey (red) lines show very large parameter biases if the assumed slope is significantly too low or too high, and in some cases the fully-modelled case fares extremely badly.  Inspection of the lower right figure shows that the systematic bias on the Dark Energy equation of state can far exceed the statistical error in the nulled analysis.  Interestingly, this panel shows that nulling is very effective in this case, showing a negligible systematic bias, even of the galaxy bias evolution assumptions are very wrong.    This insensitivity to the parameterisation of the galaxy bias evolution seems to be a fairly general feature of nulling, and is its major strength.

In the case of negligible photometric redshift errors it is sufficient to remove redshift auto-correlations from the analysis in order to avoid contributions from the $\delta \delta$ term. For a finite photometric redshift uncertainty however, $\delta \delta$ is non-vanishing also for cross-correlations between neighbouring redshift bins, and we include this effect in our study. To reduce this further potential source of parameter bias, we adopt the procedure proposed by \cite{JS09} who downweight correlations between adjacent bins in the transformed power spectra with a Gaussian centred on the mean redshift of the foreground redshift bin, choosing a width that is determined by the photometric redshift scatter (see \cite{JS09} for details).

In Table \ref{tab:magnullresults_ph-0.03} we see the effect of a more realistic photometric redshift error of $0.03(1+z)$, which is the goal for Euclid.  We still find that nulling introduces negligible cosmological parameter biases, but the statistical errors are increased, considerably in some cases.  Perhaps only for the Dark Energy equation of state parameter is there a strong case for preferring nulling over full modelling, for the galaxy evolution model considered here. 

\begin{table*}
\begin{minipage}{120mm}
\centering
\caption{Like Table \ref{tab:magnullresults}, but for a Gaussian photometric redshift scatter with width $0.03(1+z)$.}
\begin{tabular}[t]{c|cc|cc|cc|cc}
\hline\hline
parameter &  \multicolumn{2}{c|}{Version 1; nulled} & \multicolumn{2}{c|}{Version 2; standard} & \multicolumn{2}{c}{Version 2; nulled}\\
 &  $\sigma$ & $b$ & $\sigma$ & $b$ & $\sigma$ & $b$\\ 
\hline
$\Omega_{\rm m}$ &   0.426 &  0.080 & 0.012 & -0.003 & 0.136 & -0.063 \\
$\sigma_8$       &  0.468 & -0.057 & 0.009 &  0.018 & 0.165 &  0.097 \\
$h$              & 16.788 & -0.986 & 0.272 &  0.584 & 5.982 & -0.148 \\
$n_{\rm s}$       &   3.295 &  0.079 & 0.058 & -0.202 & 1.241 &  0.019 \\
$\Omega_{\rm b}$  &  1.992 & -0.091 & 0.034 &  0.046 & 0.694 & -0.042 \\
$w_0$            &   4.338 &  0.246 & 0.087 &  0.211 & 0.260 &  0.023 \\
\hline
\end{tabular}
\label{tab:magnullresults_ph-0.03}
\end{minipage}
\end{table*}

\begin{figure*}
\begin{minipage}[c]{0.5\textwidth}
\centering
\includegraphics[scale=.35,angle=0]{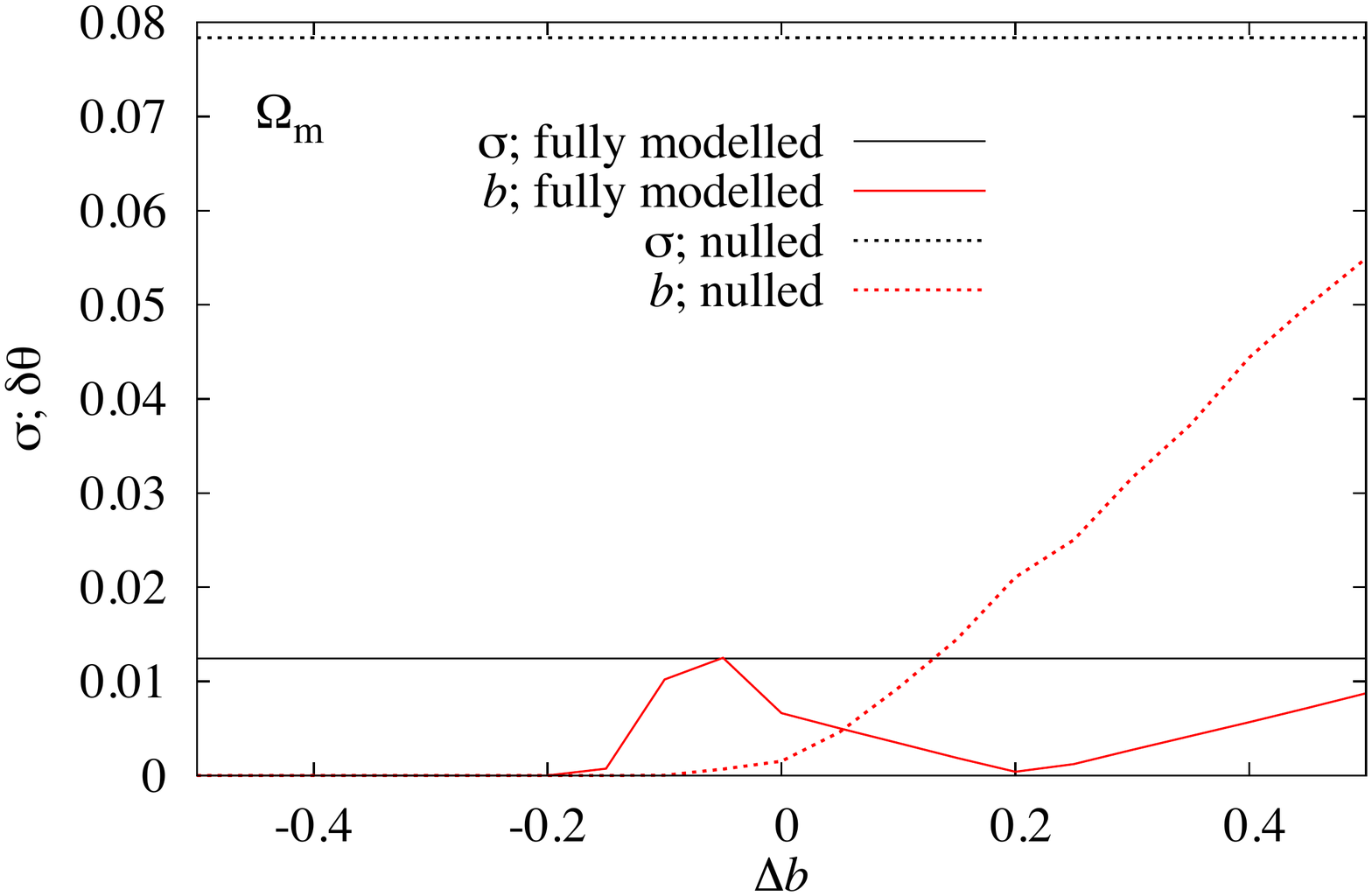}
\includegraphics[scale=.35,angle=0]{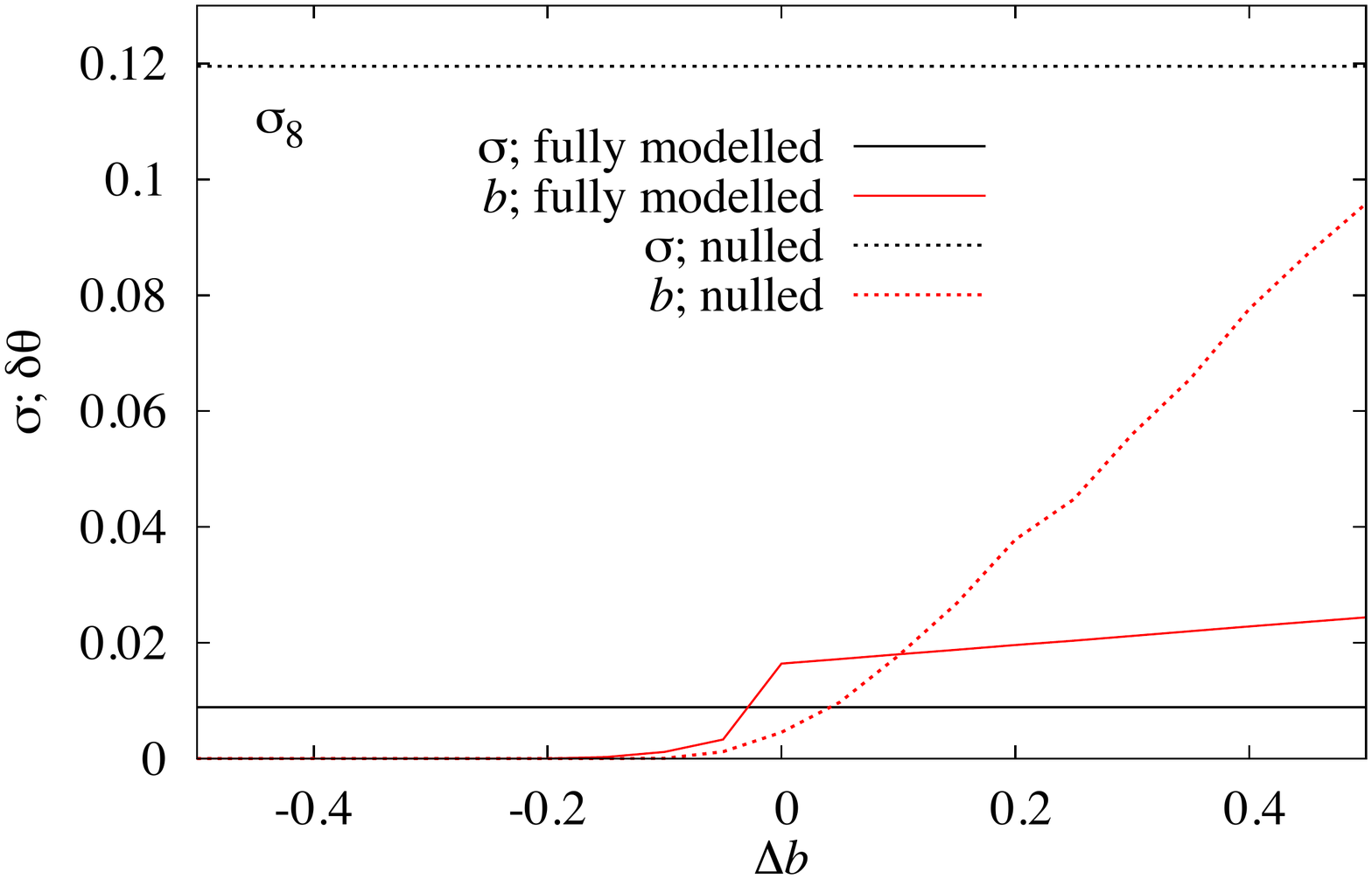}
\end{minipage}%
\begin{minipage}[c]{0.5\textwidth}
\centering
\includegraphics[scale=.35,angle=0]{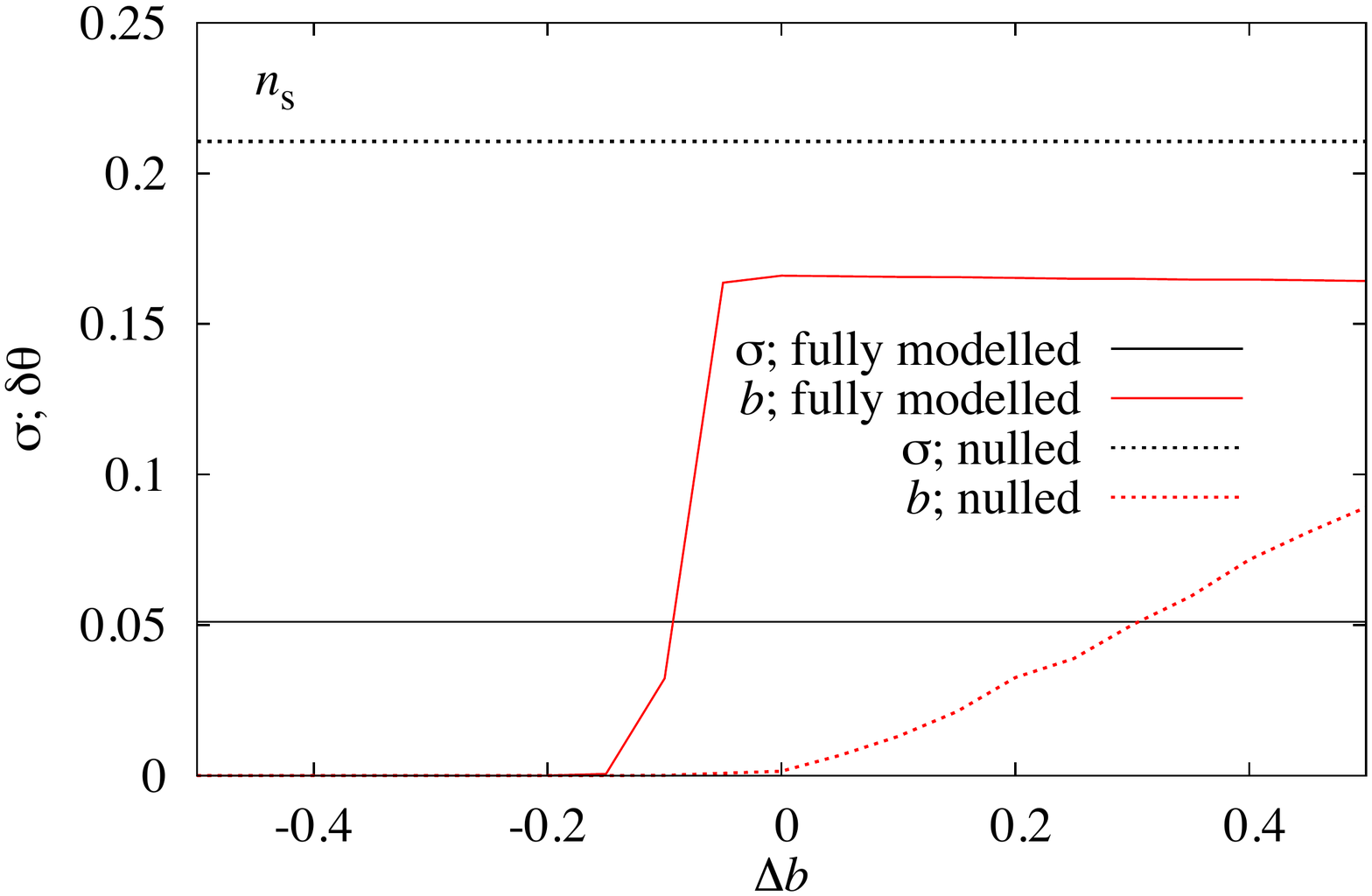}
\includegraphics[scale=.35,angle=0]{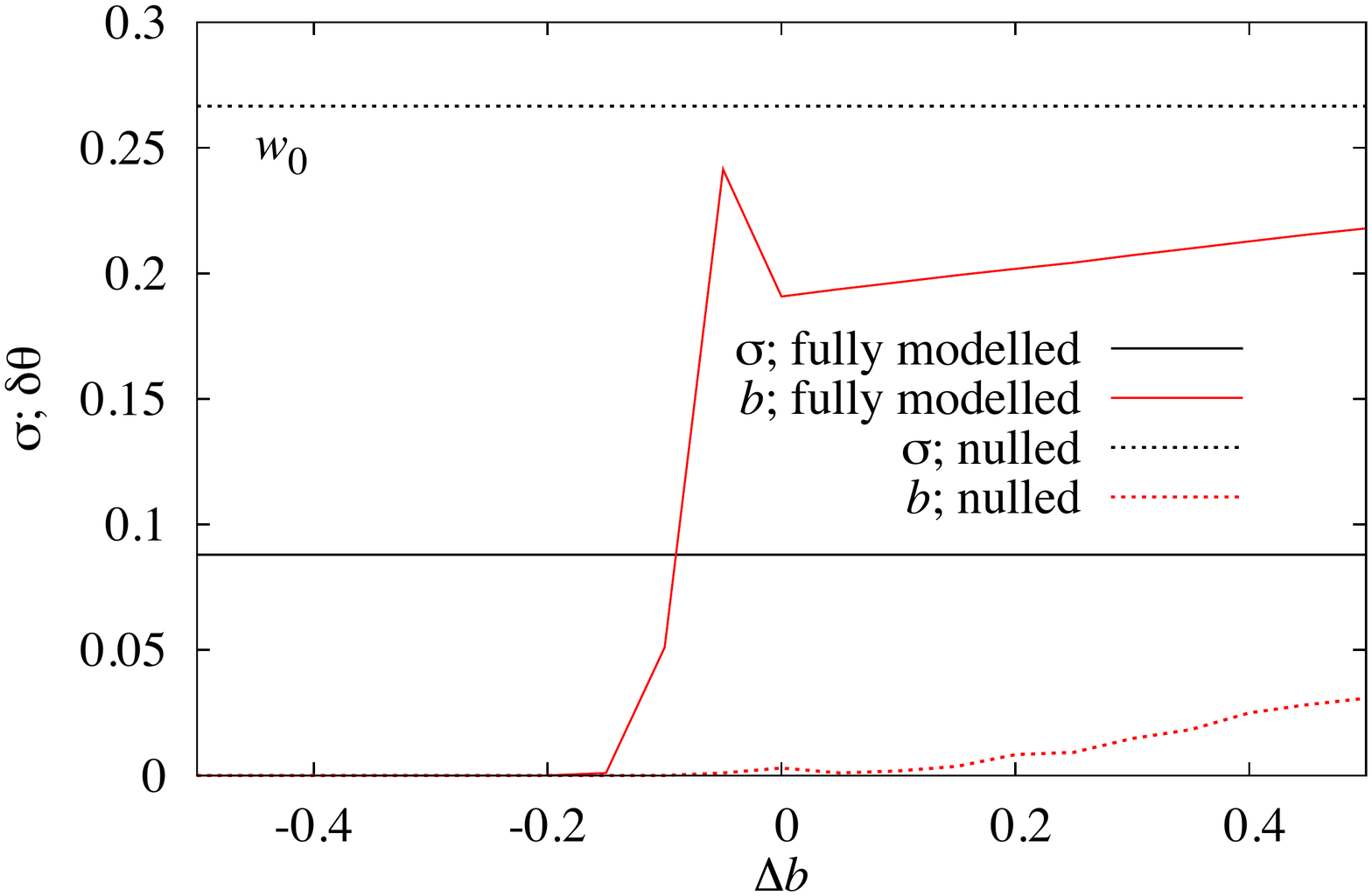}
\end{minipage}
\caption{Statistical $1\sigma$ errors (black lines) and biases (grey/red lines) for different cosmological parameters as a function of the offset $\Delta b$ of the galaxy bias model, keeping all other parameters at their fiducial values. Solid curves correspond to the results for standard analysis including marginalisation over the galaxy bias, dotted curves to the combined nulling and galaxy bias marginalisation approach.}
\label{fig:stats_offset}
\end{figure*}

In Fig. \ref{fig:stats_offset} we see the effect of varying the zero-redshift galaxy bias, $b_0$.  The true $b_0$ is $1+\Delta b$, where $\Delta b$ is the offset shown in the figure.  Recall that the bias is assumed to be constant, with a gaussian prior of width 0.1 centred on $b_0=1$, and the true bias is $1+0.1z+\Delta b$, so the average bias for galaxies is larger than $1+\Delta b$.   Hence for $\Delta b>0$, the model finds it hard to match the average galaxy bias for the lensed galaxies, but for $\Delta b<0$, the average bias is closer to unity, and the prior is better matched.  As a consequence, it is not surprising that the modelling leads to quite large biases in the parameter estimates for $\Delta b>0$.

\begin{figure*}
\begin{minipage}[c]{0.5\textwidth}
\centering
\includegraphics[scale=.35,angle=0]{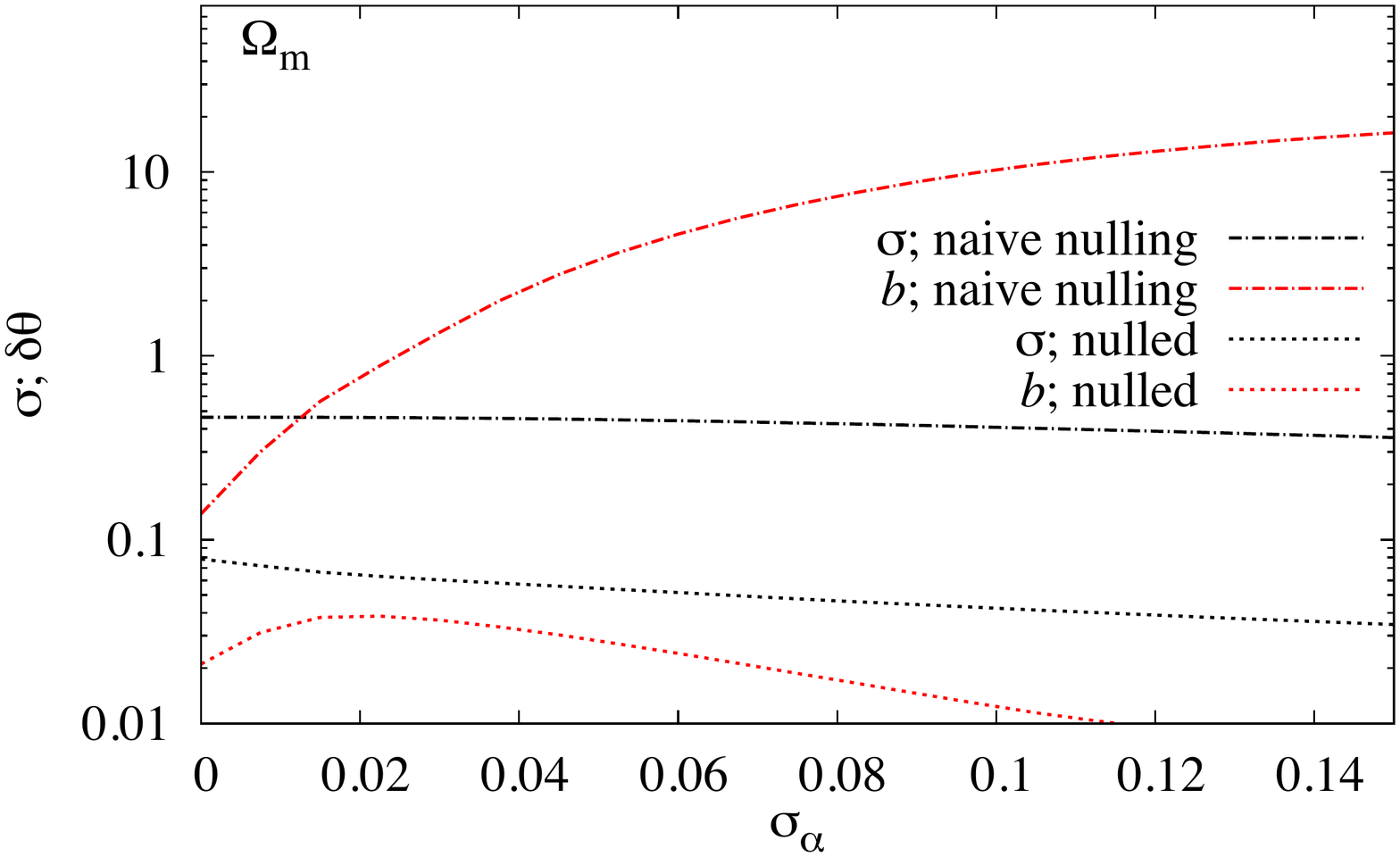}
\includegraphics[scale=.35,angle=0]{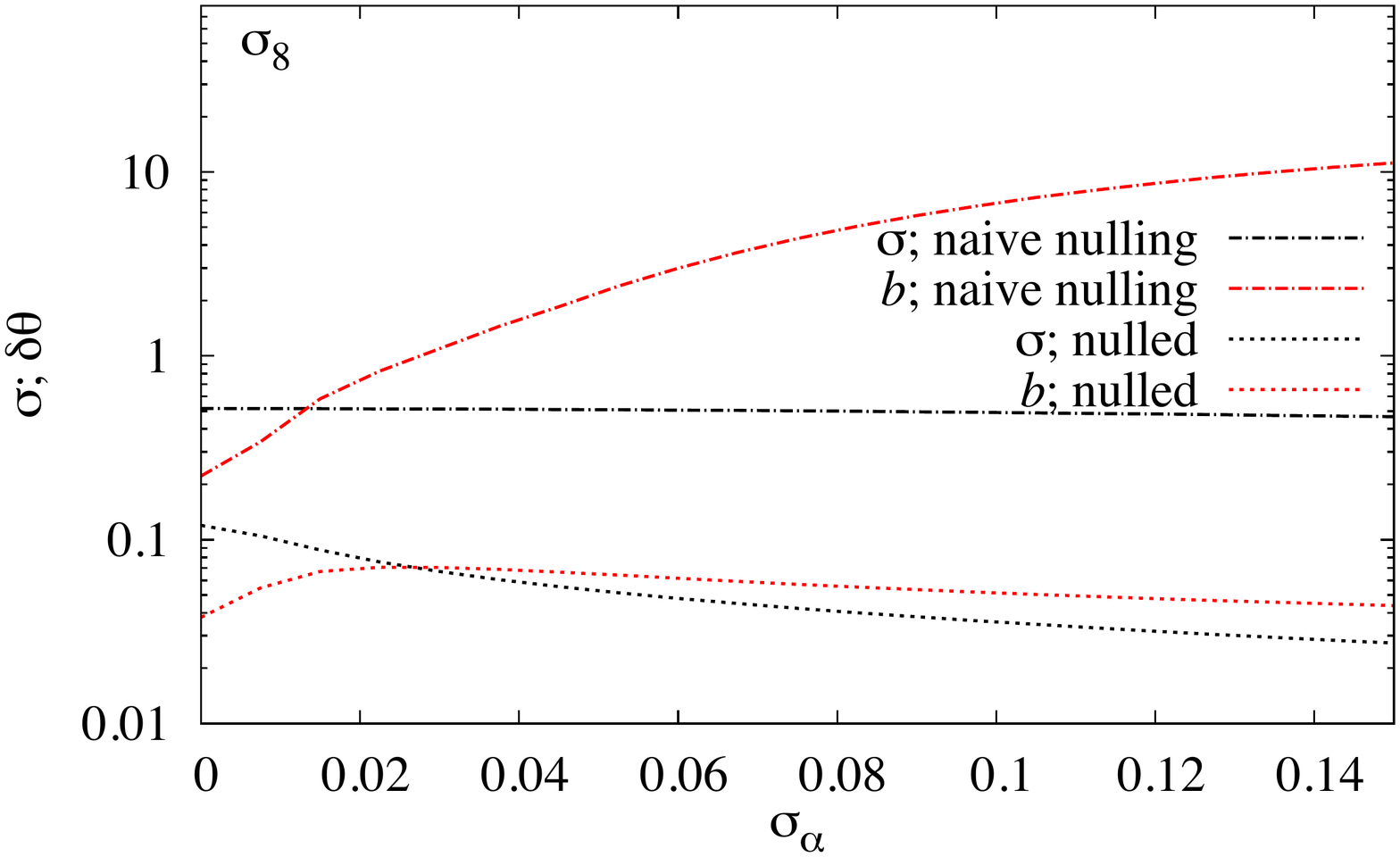}
\end{minipage}%
\begin{minipage}[c]{0.5\textwidth}
\centering
\includegraphics[scale=.35,angle=0]{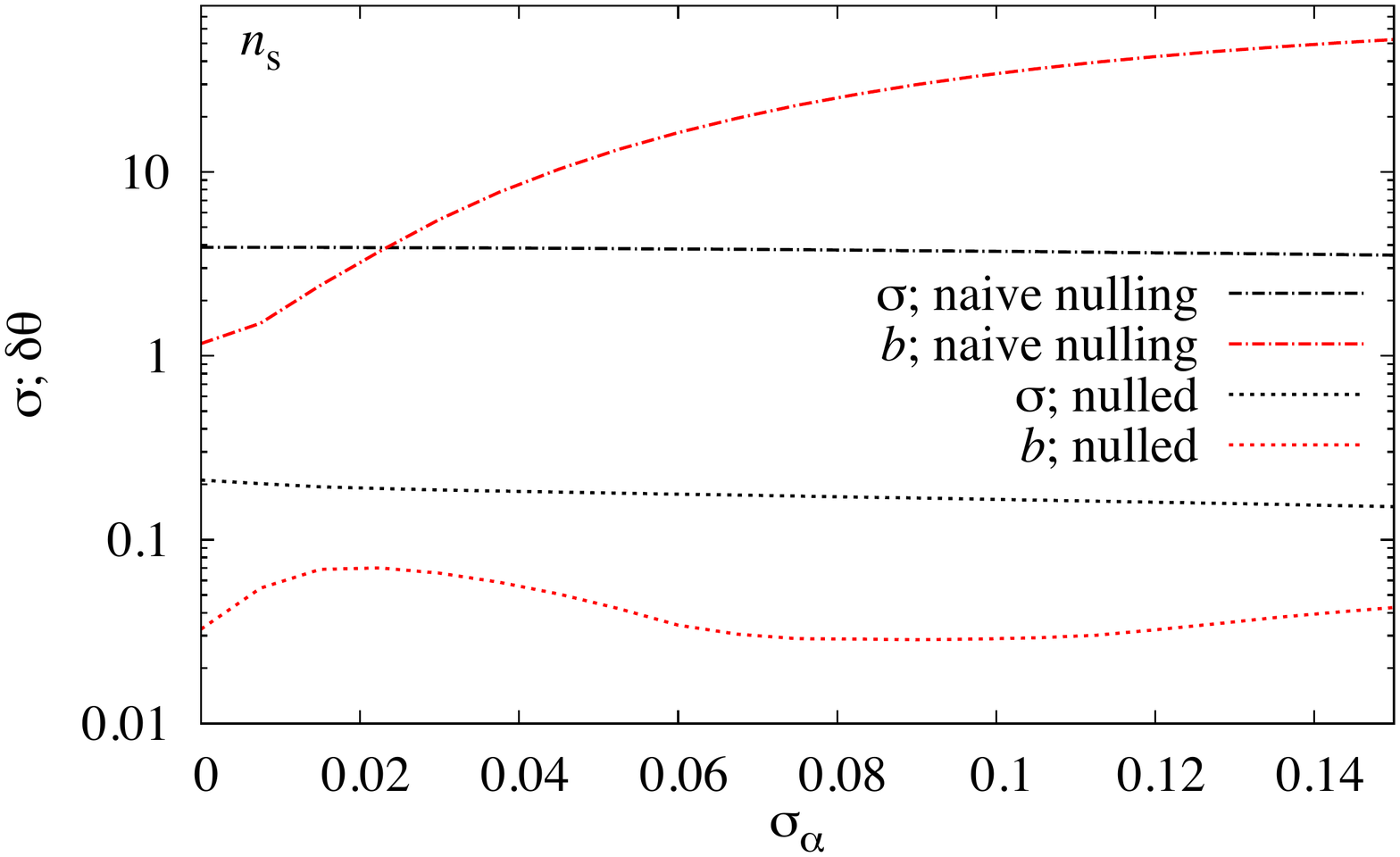}
\includegraphics[scale=.35,angle=0]{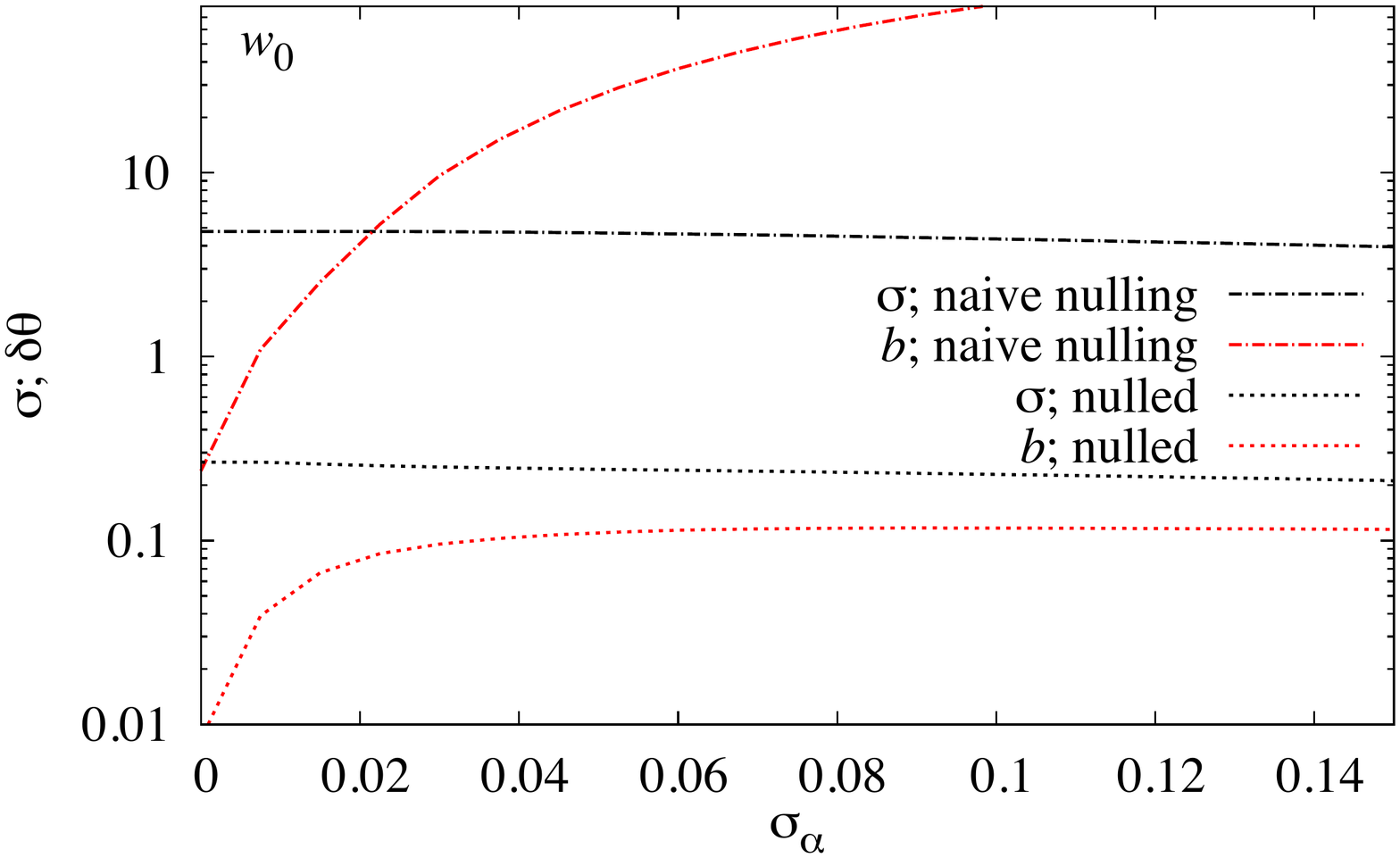}
\end{minipage}
\caption{Statistical $1\sigma$ errors (black lines) and biases (red/grey lines) for different cosmological parameters as a function of the $1\sigma$ scatter $\sigma_\alpha$ of the values of $\alpha$ in the different redshift bins, keeping all other parameters at their fiducial values. Chain-dotted curves correspond to the results for na\i ve nulling (without marginalisation over the galaxy bias), dotted curves to the combined nulling and galaxy bias marginalisation approach. The values of $\alpha$ are drawn from a Gaussian distribution centred on the fiducial values of $\alpha$ and with a scatter $\sigma_\alpha$. The results plotted are averaged over 10 realisations. To avoid noise, the same seed is used for each value of $\sigma_\alpha$.}
\label{fig:stats_alphas}
\end{figure*}

In the analysis so far, we have assumed that we know the slope of the number counts exactly.  In Fig. \ref{fig:stats_alphas} we show how sensitive the errors,  both statistical and systematic, are to errors in the number count slope.  We assume there is a gaussian random scatter with width $\sigma_\alpha$ about the true value of $\alpha$ in each redshift bin.  Note that these panels have one logarithmic axis.  We see that if we null and assume that there is no residual $\kappa\delta$ term (the `naive nulling' curves), then scatter around the fiducial number count slope is very serious for parameter errors, if the scatter exceeds about 0.02.  For nulling where we also model the contaminating $\kappa\delta$ term, the situation is much better, with parameter estimation after marginalisation over the galaxy bias being rather insensitive to variations of the slope.  In the worst cases, the parameter bias is of order the systematic error, and if the slope can be constrained to an error of less than 0.02, the systematic error is subdominant for all parameters.  Note that the rather strange behaviour of the bias in the spectral index $n_s$ for the nulled case is due to random fluctuations from a relatively small number of trials, but it is in any case subdominant for all values of the scatter probed.

\section{Discussion}

Observed galaxy clustering for deep surveys is determined by a combination of intrinsic clustering and changes in number density due to the magnification and amplification effects of galaxy lensing.   If the lensing effect could be isolated, it could be a useful cosmological probe, as it is independent of galaxy bias, and depends rather cleanly on cosmological parameters.   The difficulty is that the signal is small in comparison with the intrinsic clustering term.  In this paper, we have presented a method to isolate as far as possible the part of the clustering signal which is due to cosmic magnification.  This involves exploiting the known redshift dependence of the magnification signal to remove the cross-terms which link the magnification with foreground intrinsic clustering.   By isolating the magnification term, we aim to remove the dependence of clustering on the galaxy bias, which is subject to uncertainties connected with galaxy formation efficiency, leaving a clean test of cosmology relying only on the behaviour of the matter power spectrum and distance-redshift relations, both of which are much more easily predictable theoretically than galaxy bias.  This is a very challenging task, because the contamination of the clean signal by intrinsic clustering is very large, and we have been partially successful.  The intrinsic clustering signal can be reduced very substantially by the `nulling' technique we have employed (by a factor of about 100), but not to a negligible level, so the residual contamination still has to be modelled.  However, since the amplitude of the residual contamination is much smaller than the signal before nulling, we do not need to model bias as accurately.    By investigating an example cosmological setup where galaxy bias has a mild evolution with redshift, and modelling it incorrectly as a constant bias, we find that full modelling of the clustering signal (including the intrinsic clustering term) has small statistical errors, principally because the $\kappa\delta$ term is carrying quite a lot of cosmological information, but the incorrect parametrisation of the galaxy bias leads to systematic errors in the cosmological parameters which are typically significantly larger than the statistical errors.  Nulling, on the other hand, reduces the sensitivity to the galaxy bias modelling, as most of the effect of galaxy bias is removed.  The result is that the systematic errors in the cosmological parameter estimates are reduced to a low level, at the expense of  increases in statistical errors, which in some cases are quite large.  The reason for the increase in error bars is that the $\kappa\delta$ term is relatively large (in contrast to the GI term in cosmic shear), and it contains cosmological information, so removing it does increase the errors.  Thus nulling gives a robust alternative analysis to modelling the full clustering signal, as it is less sensitive to modelling of the galaxy bias, but it is a conservative analysis.  In short, full modelling is subject to a bias of a size which we may not know, as it depends on how wrong our modelling of galaxy bias is. Nulling protects us against this.
However,  the realistic position is not so pessimistic, as we do in principle have empirical measurements of bias available, from the auto-correlation of number counts in tomographic bins, or from the galaxy bispectrum, which may prove helpful.

One interesting technical point is how we are able to improve our parameter estimation by throwing away data.  We see that the weighted cross-powers are simply linear combinations of individual bin-bin cross-power, but we have simply removed one linear combination of the bin number densities.   The reason is as follows.  The removed combination has a very large contribution from intrinsic clustering, and is thus sensitive to how the galaxy bias is modelled.  If this is done imperfectly, for example by assuming some redshift dependence which does not hold in nature, then the parameter estimation process may try to fit this combination, at the expense of a poorer fit to the nulled combinations.  As a result the parameter estimation may be systematically wrong.  By nulling the intrinsic signal, the cross-power spectra have a much greater signal coming from magnification, independent of bias, so the fitting is more likely to find the correct parameters.  Note that if we know the bias evolution precisely, then it will always be better to model all the cross-power, as this does not lose information.  In practical cases, it is an open question which method is preferred, and this will be the subject of a further paper.

The key issue for the use of magnification for cosmological parameter estimation will be the level of systematic errors, since surveys which are beginning now and which are planned for the future are large enough that it is likely that the statistical errors will become irrelevant.  We have not addressed this issue in the paper, except to an extent discussing the systematic errors arising from imperfect bias modelling.  Evidently, zero-point magnitude errors will feed into the error budget, although recent improvements in photometry (e.g. \citealt{Ilbert06}) are encouraging.  Note that the principal effect here will be angular on the sky, so the redshift dependence of the lensing signal may help to alleviate this issue.  Perhaps of most concern is the photometric redshift estimation (e.g. \citealt{Hil10} and references therein), where catastrophic outliers will correlate intrinsic clustering over large redshift separations, and mimic the magnification signal to some degree, and the nulling process itself is also sensitive to outliers (see \citealt{JS09} for details). Both of these require further study to see what requirements will be needed to be met for cosmic magnification to be a viable technique.

\section*{Acknowledgments}

\appendix

\bsp

\label{lastpage}

\end{document}